# Fe₃Se₄: A Possible Ferrimagnetic Half-Metal?

Girish C. Tewari[1*&], Divya Srivastava[1&], Reijo Pohjonen[2], Otto Mustonen[1,3], Antti J. Karttunen[1], Johan Lindén[2] and Maarit Karppinen[1]

[1]*Department of Chemistry and Materials Science, Aalto University, FI-00076 Aalto, Finland*
[2]*Faculty of Natural Sciences and Engineering/Physics, Åbo Akademi University, FI-20500 Turku, Finland*
[3]*Department of Materials Science and Engineering, University of Sheffield, Sheffield, S1 3JD, United Kingdom*
(Dated: 05/21/2020)

Half-metallic ferromagnets show 100% spin-polarization at the Fermi level and are ideal candidates for spintronic applications. Despite the extensive research in the field, very few materials have been discovered so far. Here we present results of electronic band structure calculations based on density functional theory and extensive physical-property measurements for Fe₃Se₄ revealing signatures of half-metallicity. The spin-polarized electronic band structure calculations predict half-metallic ferrimagnetism for Fe₃Se₄. The electrical resistivity follows exponentially suppressed electron-magnon scattering mechanism in the low-temperature regime and show a magnetoresistance effect that changes the sign from negative to positive with decreasing temperature around 100 K. Other intriguing observations include the anomalous behavior of Hall resistance below 100 K and an anomalous Hall coefficient that roughly follows the $\rho^2$ behavior.

Highly spin-polarized ferromagnetic (FM) metals and semiconductors are exciting candidates for spintronics devices [1,2]. Depending on their electronic band structure at the Fermi level ($E_F$), these systems can be divided into three categories: (i) spin-gapless FM semiconductors, (ii) half-metallic (HM) FMs, and (iii) zero-gap HM FMs. This classification is based on the way the conduction and valence band edges of the majority spin states are connected at the $E_F$, i.e. gapped, overlapping or touching, while the minority spin state edge bands are gapped [3,4].

In HM FMs, the complete absence of minority spin states at $E_F$ implies the existence of an energy gap for only minority spin states, and leads to exceptional electronic transport properties such as an anomalous Hall effect (AHE) and a crossover from positive to negative magnetoresistance (MR) effect due to the competition between spin scattering and other scatterings of charge carriers, in the presence of external magnetic field [4–7]. In particular, the quantum phenomenon AHE is currently attracting considerable attention [7–12]; it is a sum of intrinsic and extrinsic (skew-scattering and side-jump) contributions. The intrinsic contribution depends on the band structure, related to the Berry phase and Berry curvature in the momentum space, while the extrinsic contribution is related to sample magnetization. Apart from this, electrical resistivity follows an exponentially suppressed $T^2$-dependence ($\rho(T) = AT^2 e^{\left(-\frac{\Delta}{T}\right)}$) at temperatures below the energy-gap temperature ($\Delta$), which in typical FMs originates from electron-magnon scattering [12–15]. This happens due to the suppression of spin-flipping of conduction electrons in the presence of a gap for minority spin states. Experimentally such behavior has been observed in very few HM materials such as granular thin films of CrO₂ [15], and the Heusler alloys, Co₂FeSi [7] and (Co,Ti)FeGe [16].

Theoretically, many Heusler alloys in particular have been predicted to show the HM FM behavior [17]. There are also numerous experimental studies aiming at finding new HM FMs [18–24]; nonetheless, very few promising new candidates have been discovered. In this paper, we report low-temperature electronic transport and magnetotransport properties and spin-polarized electronic band structure calculation results for an iron-selenide compound Fe₃Se₄ that strongly indicate towards a ferrimagnetic (FiM) HM.

The iron selenide phases FeSe_x ($1 \leq x \leq 1.33$) with excess selenium crystallize in a monoclinic crystal structure, derived from the hexagonal NiAs-type structure [25,26]. The non-stoichiometry leads to the formation and ordering of Fe vacancies; among these phases, Fe₃Se₄ [27] and Fe₇Se₈ [28] are known to be ferrimagnetic. The FiM structure originates from (i) ordering of the Fe vacancies into every second iron-atom (Fe1) plane, (ii) FM alignment of moments within both the partially-vacant (Fe1) and the fully-occupied (Fe2) iron-atom planes, and (iii) antiferromagnetic (AFM) coupling between neighboring planes [25,27]. Early electronic transport measurements in the temperature range of 100 to 1000 K on single-crystal Fe₃Se₄ and Fe₇Se₈ samples revealed metallic and semiconducting behaviors below and above the Curie temperature ($T_C$=330 K), respectively [27,28]. A more recent study on Fe₃Se₄ nanowire arrays in a porous anodic aluminum oxide membrane reported semiconducting behavior [29]. The zero-field resistivity followed variable-range hopping (VRH) mechanism below 295 K, and the small positive MR effect observed below 100 K was attributed to the effect of magnetic field on the VRH conduction. However, in these previous studies, the low temperature transport properties of $Fe_3Se_4$ were not investigated in detail, in particular below 100 K. Apart from this, the effect of Se stoichiometry on the transport properties was not studied either. Here we present electronic transport and magnetotransport properties of ferrimagnetic $Fe_3Se_4$ in the low temperature regime from 2 to 300 K. We also discuss the effect of Se stoichiometry on the transport properties.

*girish.tewari@aalto.fi; Corresponding author
&Authors contributed equally

Two samples were prepared through solid-state synthesis: slightly non-stoichiometric $FeSe_{1.28}$ and stoichiometric $Fe_3Se_4$ ($FeSe_{1.33}$). Appropriate amounts of Fe (99.99%) powder and Se (99.99%) shots were homogenized, pressed into a pellet at 10 kbar, vacuum sealed in a quartz tube, slowly heated up to 750 °C and held for 48 h. The reacted charge was reground, pressed again into a pellet and annealed at 400 °C for 18 h and cooled slowly to room temperature [30]. X-ray diffraction (XRD; PANanalytical X'Pert PRO MPD Alpha-1; Cu Kα1 radiation) patterns for the two polycrystalline samples are shown in Fig. 1. The two patterns are essentially identical, and both could be readily indexed with the monoclinic space group $I2/m$ (#12), indicating phase purity. Rietveld analysis of the non-stochiometric $FeSe_{1.28}$ sample carried out in the $I2/m$ space group was published in our previous paper [30]. For $FeSe_{1.33}$, the peaks are slightly shifted to the higher angles, indicating tiny lattice contraction with increasing Se content.

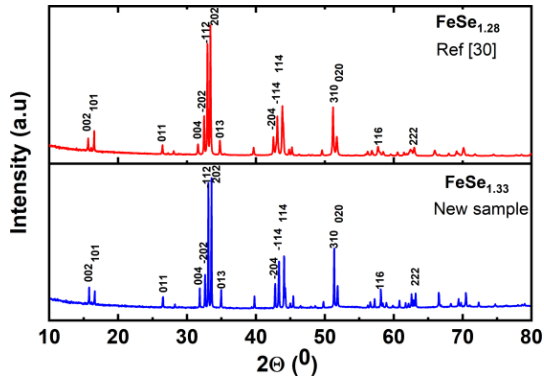

FIG. 1. X-ray diffraction patterns of slightly non-stoichiometric ($FeSe_{1.28}$) [30] and stoichiometric ($FeSe_{1.33}$).

The electronic transport measurements were performed on $FeSe_{1.28}$ using a Physical Property Measurement System (PPMS; Quantum Design) device equipped with a 9 T magnet. Electrical resistivity ($\rho$) and Hall effect were measured using the standard four-point-probe technique. Magnetization versus magnetic field ($M-H$) isotherms were collected by sweeping the magnetic field from -5 to 5 T using a vibrating sample magnetometer (VSM; Quantum Design PPMS).

Zero-temperature structural optimization and electronic structure calculations were conducted in the frame work of spin-polarized density functional theory (DFT) using the CRYSTAL17 [31] software package. The HSE06 [32] hybrid functional was used to describe the exchange and correlation. The DFT calculations were all-electron calculations using a triple-ζ-valence + polarization (TZVP) level Gaussian-type basis set [33]. Valence and core orbitals were described by three and one atom-centered Gaussian basis functions, respectively. Coulomb and exchange integral tolerance factors (TOLINTEG) were set to tight values of 8, 8, 8, 8, and 16. In the geometry optimization, the reciprocal space was sampled using an 8×8×8 *k*-grid [34].

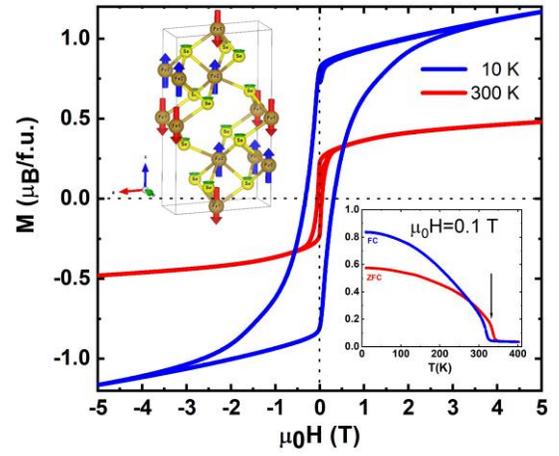

FIG. 2. Magnetization versus magnetic field measured at 10 and 300 K. Bottom inset: $M$ versus $T$ measured under field-cooled (FC) and zero-field-cooled (ZFC) conditions. The arrow indicates the Curie temperature. Top inset: Optimized magnetic structure of $Fe_3Se_4$ showing orientation and magnitude of moments on the Fe1 (red), Fe2 (blue) and Se (green) atoms.

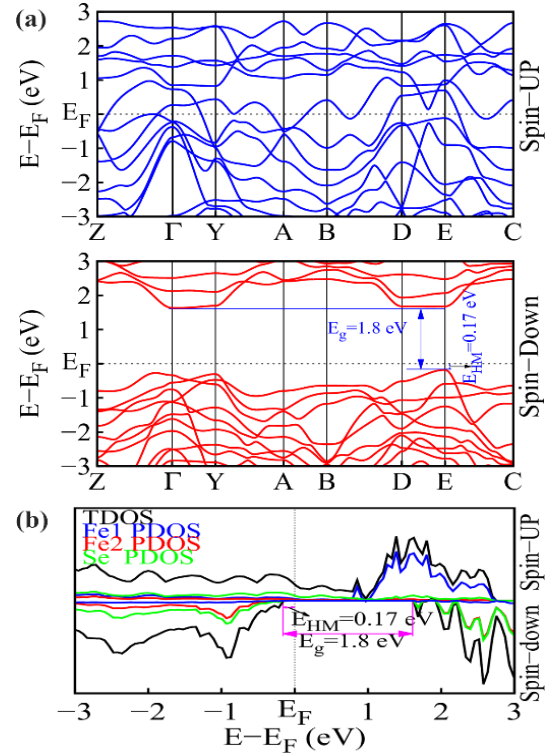

FIG. 3. (a) DFT Calculated spin resolved electronic band structure, and (b) total and atom resolved partial density of states for FiM $Fe_3Se_4$; $E_g$ and $E_{HM}$ correspond to electronic and half-metallic energy gap, respectively.

The unit cell of $Fe_3Se_4$ was optimized in monoclinic space group *I2/m* [25,30]. In structural optimizations, both the lattice parameters and atomic positions were fully relaxed. Our optimized lattice parameters, *a*=6.45 Å, *b*=3.67 Å, *c*=11.74 Å and *β*=92.84⁰, are somewhat larger than the experimental values of *a*=6.20 Å, *b*=3.53 Å, *c*=11.38 Å and *β*=91.83 [25,30]. In the DFT-HSE06 optimized crystal structure of FiM $Fe_3Se_4$ (top inset of Figure 1) Fe atoms occupy two non-equivalent positions: atoms located in the layers containing vacancies are labelled as Fe1 and those in the fully-occupied layers are labeled as Fe2. The spin density is mainly located on the Fe atoms, the magnetic moments being -3.79 $\mu_B$ for Fe1 and 3.52 $\mu_B$ for Fe2. However, Se atoms also possess very small magnetic moments compared to Fe atoms which varies from 0.012 to 0.14$\mu_B$.

2.00 $\mu_B$/f.u., in line with the previously DFT-PBE calculated values of 2.12–2.17 $\mu_B$/f.u. [35,36], but nearly doubling the reported experimental value of 1.17 $\mu_B$/f.u. at 10 K [25,27]. Our experimental values are 0.48 $\mu_B$/f.u. at 300 K and 1.17 $\mu_B$/f.u (0.39 $\mu_B$/atom) at 10 K, which are well in the range of previously reported experimental values [25,27]. The large discrepancy in magnetic moment between experimental value and DFT predicted value indicate delocalization of 3d electrons of iron ions and strong interactions are present between iron ions and their neighbors.

The spin-polarized electronic band structure and atom-projected density of states (DOS) of FiM $Fe_3Se_4$ are depicted in Figure 3; the result points toward HM ferrimagnet. The coordinates of the high-symmetric k-points used for plotting band structure are (0, 0, 0.5), (0, 0, 0), (0, 0.5 ,0), (-0.5, 0.5, 0), (-0.5, 0, 0), (-0.5, 0.5, 0.5), (-0.5, 0.5 , 0.5), (0, 0.5, 0.5) for Z, Γ, Y, A, B, D, E and C, respectively. The spin-up channel is metallic with hole and electron pockets, whereas the spin-down channel is semiconducting with an indirect band-gap of 1.8 *eV*. The valence band maxima (VBM) occurs at E point, with conduction band minima at Γ point. The direct band-gap is 1.85 eV, occurring at E point. This implies that the spin-up electrons dominantly contribute to the charge transport, and the current is fully spin-polarized. From Figure 3(b), the main contribution to the spin-up channel at $E_F$ is from Se, with small mixing of Fe1 and Fe2 atoms. In case of the spin-down channel, there is strong coupling between Se and Fe2 at VBM and CBM. PDOS of Se gives more density of states near the Fermi level compared to those of Fe1 and Fe2, suggesting that Se makes a major contribution to the half-metallicity in $Fe_3Se_4$.

Zero-field resistivity as a function of temperature is displayed in Figure 4, showing metallic-like temperature dependence. The residual resistivity ($\rho_R$) value is 3.62 *mΩ-cm* yielding the residual resistivity ratio, RRR = $\rho$(300 K)/$\rho_R$ = 3.92. The low-temperature data were analyzed to understand the transport mechanism involved in $Fe_3Se_4$. For a FM metallic system, $\rho$ is a sum of $\rho_R$, and the magnonic ($\rho_M$) and phononic ($\rho_P$) terms:

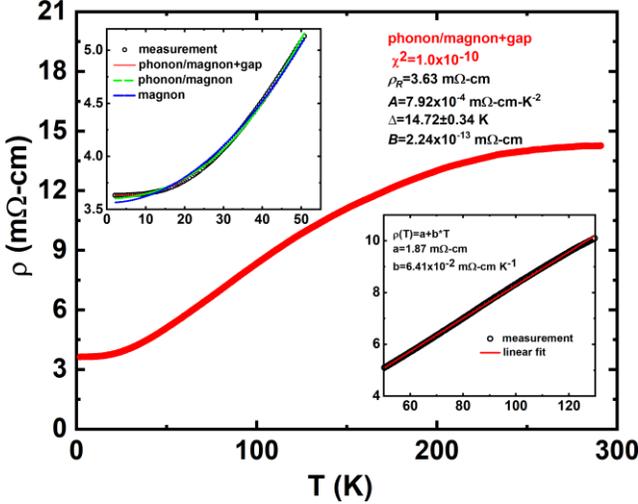

FIG. 4. Resistivity versus temperature. Upper inset: the low-T data fitted with phonon/magnon+gap, phonon/magnon and pure magnon scattering terms. Lower inset: a red line shows liner fit of the data from 50 to 120 K. The measured data is shown by black circles. Phonon/magnon+gap fitting parameters are shown.

In Figure 2, we display the *M-H* hysteresis loops for our $Fe_3Se_4$ sample measured at 10 and 300 K; the *M-T* curves in the inset reveal the $T_C$ around 340 K. None of the *M-H* curves show perfect saturation (with magnetic field) even up to the highest measurement field of 5 T. At high fields above 2 T, *M* increases linearly with *H* with a negligible slope; this behavior could be related to the FiM nature of $Fe_3Se_4$. The absence of saturation of *M* at high fields indicates flipping of minority spins along the magnetic field. From our DFT optimized FiM $Fe_3Se_4$ structures the total magnetic moment was yielded at

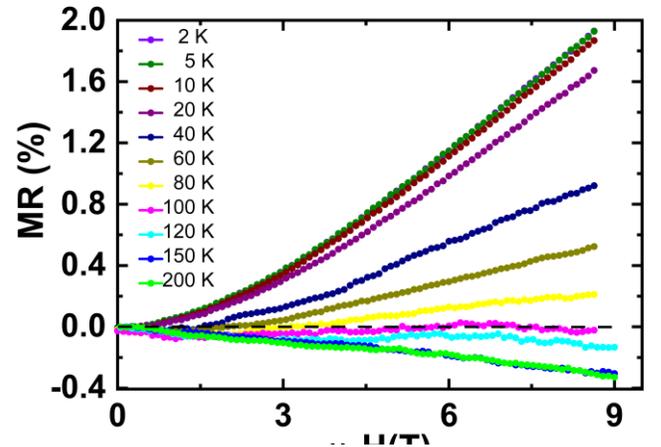

FIG. 5. Magnetic field dependence of MR measured at various temperatures.

$$\rho(T) = \rho_R + \rho_M(T) + \rho_P(T) \qquad (1)$$

Among these parameters, $\rho_R$ is temperature independent (being caused by grain boundaries and defects in polycrystalline sample). For a HM FM system, $\rho_M$ can be defined by the usual electron-magnon scattering with a Boltzmann factor as

$$\rho_M(T) = AT^2 e^{\left(-\frac{\Delta}{T}\right)} \qquad (2)$$

where $A$ is magnon scattering strength and $\Delta$ measures the energy gap between the $E_F$ and the band edge of minority spin carriers [6,14,15]. The value of $\rho_P$ can be calculated by Bloch-Grüneisen formula

$$\rho_p(T) = B \left(\frac{T}{\theta_D}\right)^5 \int_0^{\frac{\theta_D}{T}} \frac{x^5}{(e^x-1)(1-e^{-x})} dx \qquad (3)$$

where $B$ is phonon scattering exponent and $\theta_D$ is Debye temperature. For $Fe_3Se_4$, the estimated value of $\theta_D$ is 262 K, from specific heat measurement. The measured $\rho(T)$ data do not follow Eq. (1) over the entire measured temperature range. However, below 50 K, the data can be well fitted by Eq. (1), see the top inset in Figure 4. The very small value of exponent $B$ clearly shows negligible contribution of phononic term in $\rho$ below 50 K. In a HM FM, at $E_F$ the minority spin states are gapped, and the minimum energy required by majority spin carriers to occupy minority spin states through a spin flip is $k_B\Delta$. Our fitting yields the $\Delta$ value of 14.72±0.35 K and the corresponding energy gap of 1.3 *me*V. The HM energy gap ($E_{HM}$) is extremely small in comparison to the gap of 0.17 $e$V predicted through theoretical band structure calculation. The fact that $\rho(T)$ follows an exponentially suppressed spin-flip scattering at low temperatures clearly indicates towards the HM nature of $Fe_3Se_4$; a fully spin-polarized state at $E_F$ can be realized for $T < \Delta$. We also tried to fit the data with pure magnon ($\rho(T) = \rho_R + AT^2$) and magnon plus phonon scattering ($\rho(T) = \rho_R + AT^2 + \rho_p(T)$) terms. It can be clearly seen from the top inset of Figure 4, the data do not fit well neither the pure magnon nor the magnon plus phonon scatterings in comparison to phonon/magnon+gap given by Eqn. (1). This indicates that $\rho(T)$ does not follow the conventional electron-magnon scattering of a band ferromagnet. Above 50 K up to 120 K, a linear $T$-dependence is seen from the bottom inset of Figure 4. This signifies the domination of $\rho$ by $\rho_p$ at elevated temperatures. Above 120 K, $\rho(T)$ shows complex $T$-dependence. The band structure calculations support the experimental finding that $Fe_3Se_4$ is HM FiM. DFT-HSE06 revealed a much larger $E_{HM}$ of 0.17 eV compared to the experimentally estimated value of 1.3 meV. This may be due to the tiny Se deficiency of our $Fe_3Se_4$ sample, i.e. $FeSe_{1.28}$ instead of $FeSe_{1.33}$ [30]. The presence of Se deficiency in our $Fe_3Se_4$ sample may account for the difference between the calculated and the experimental HM gap. The calculated atom projected DOS (Figure 3) shows that contribution of Se atoms near the Fermi level are more compared to Fe atoms. Therefore, vacancy defects could give more states near the Fermi level both in VB and CB, hence reducing the gap. To confirm this, we investigated the stoichiometric $FeSe_{1.33}$ sample. The $\rho$ versus $T$ data for this sample could be fitted with Eqn. (1) up to much higher $T$ around 230 K. The fitting resulted in much larger value of $\Delta$=395±10 K and the corresponding value of HM gap is found to be around 34 meV. This clearly shows that the HM gap is strongly affected by Se stoichiometry.

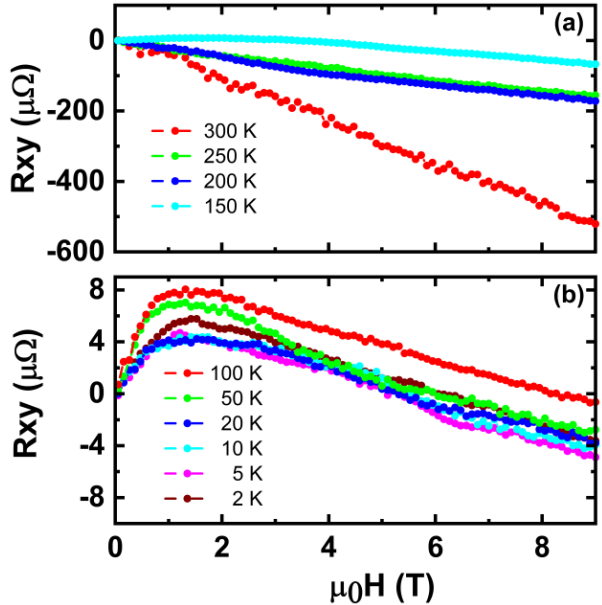

FIG. 6. Hall resistance versus magnetic field measured at various temperatures.

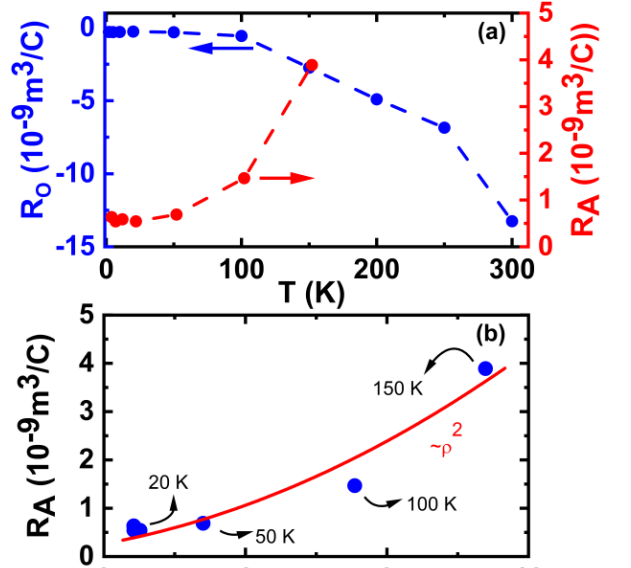

FIG. 7. (a) Ordinary and anomalous Hall coefficient versus temperature. (b) $R_A$ versus $\rho$ (blue circles); red line is the quadratic fit.

In Figure 5, we present the measured MR ratio (= $100[\rho(H) - \rho(0)]/\rho(0)$) against the applied magnetic field

($H$) at various temperatures. The sign of MR is positive at low temperatures and negative at high temperatures; the crossover happens around 100 K. The magnitude of the positive 9-T MR at 2 K is ca. 1.8% and the negative 9-T MR at 200 K is ca. 0.3%. The negative MR at high temperatures is related to the suppression of spin-dependent scattering in the presence of $H$. The positive MR normally arises due to the Lorentz force on charge carriers; accordingly the MR should show quadratic $H$ dependence [3]. Here, however, the positive MR for our $Fe_3Se_4$ sample shows a $H$ dependence between $H^{1.5}$ and $H$ (in the low and high field, respectively), ruling out this possibility. On the other hand, if we consider the HM FiM behavior of $Fe_3Se_4$, at low-$T$ spin scattering is suppressed exponentially due to the gapped minority spin channel at $E_F$ and the scattering through defects dominates and may give rise to the positive MR, characteristics of a multiband metal [3,4,7]. Previous experimental studies on half metallic $Co_2FeSi$ [7] and spin gapless semi-conductors $FeCoSi$ [3], $Mn_2CoAl$ [4] also showed a crossover from negative to positive in MR with lowering $T$. The MR was found to be nonsaturating and nearly linear with $H$ at high $H$. The positive MR in case of $Co_2FeSi$ was attributed to its half-metallic nature and expected to freeze out electron-magnon scattering exponentially at low temperature. Hence, electron-defect scattering must become important and expected to show a positive MR [7]. On the other hand, in case of spin gapless semi-conductors, the quantum linear MR was originated from linear energy spectrum and the participation of only one Landau band in the conductivity [5].

We now discuss the anomalous behavior of Hall effect in $Fe_3Se_4$. Hall resistance ($R_{xy}$) in magnetic materials arises from two effects: (i) normal scattering of charge carriers due to Lorentz force in the presence of external magnetic field perpendicular to the applied current, and (ii) anomalous scattering due to internal magnetic field arising from the spontaneous magnetization of FM/FiM materials. These two effects result, respectively, in the linear dependence of $R_{xy}$ on the applied magnetic field along with a sharp change in slope around the saturation magnetic field.

In Figure 6, we display the $R_{xy}$ data for $Fe_3Se_4$; the upper panel shows the high-$T$ behavior of $R_{xy}$, where the normal linear dependence with a negative slope is observed; then upon cooling, $R_{xy}$ changes its behavior below 150 K, as seen from the lower panel. Below 150 K, $R_{xy}$ increases linearly with magnetic field, while above the crossover field around 1.5 T it rather decreases linearly. The crossover field corresponds to the changeover seen in the magnetization behavior (Figure 2). In a magnetic system, $R_{xy}$ depends on the ordinary ($R_O$) and anomalous ($R_A$) Hall coefficients as follows: $R_{xy} = \mu_0(R_O H + R_A M)$. We estimated the value of $R_O$ from the slope of $R_{xy}$ vs $H$ at high fields, and then by using $R_O$ and the value of $M$ at the crossover magnetic field, the value of $R_A$ was estimated from the low field data; in Figure 7(a), we display the $T$ dependence of $R_O$ and $R_A$. For $T < 100$ K, $R_O$ is small and remains almost constant, typical of a metal. Above 100 K it rapidly increases, presumably due to a decrease in charge carrier mobility. Below 50 K, $R_A$ is of the order of magnitude as $R_O$, and independent of $T$, but rises strongly with $T$ like $R_O$ above 50 K. Above 150 K the anomalous behavior disappears. At 150 K, $R_A$ ($4 \times 10^{-9} m^3$/C) is larger than $R_O$ ($2.5 \times 10^{-9} m^3$/C). To explain the mechanism behind the anomalous behavior of Hall resistance, we plot $R_A$ with zero-field $\rho$ for various $T$'s (blue circles in Figure 7(b)); the quadratic dependence of $R_A$ on $\rho$ is shown by a solid red line. Also, the anomalous Hall conductivity (AHC; $\sigma_{xy} = \rho_{xy} / \rho^2_{xx}$, where $\rho_{xx}$ = zero field resistivity, $\rho_{xy}$=Hall resistivity) derived from the measurements was found to have an extremely low value of 0.014 S/cm at 2 K, changing then slowly with $T$ to 0.008 S/cm at 150 K. We speculate that it might be related the net Berry curvature and related to the topology of electronic band structure of the material. The local Berry curvature in the Brillouin zone (BZ) changes sign on reversing the sign of momentum vector [4,12]. The $Fe_3Se_4$ system (space group I2/m) system holds time reversal and mirror symmetries so the net Berry curvature and thus the AHC is negligible. The negligibly small low-$T$ AHC values being nearly $T$ independent correspond to $R_A \sim \rho^2$, commonly explained as evidence for the intrinsic Berry phase and/or side-jump scattering contributions [4,7,11,12].

In conclusion, our electronic band structure calculations and experimental low-temperature electronic transport and magnetotransport measurements strongly point towards $Fe_3Se_4$ being a ferrimagnetic half-metal. From the calculated spin-polarized electronic band structure point of view, $Fe_3Se_4$ appears as a half-metallic ferrimagnet with the HM gap of 0.17 eV. However, the experimental HM energy gap for $Fe_3Se_4$ is very narrow, in the range of 1.3 to 34 meV depending on the Se stoichiometry. Experimentally, we observe a crossover from a negative to a positive MR effect with decreasing temperature. Moreover, the Hall resistance displays an anomalous behavior below 100 K, the anomalous Hall coefficient follows quadratic behavior as a function of zero-field resistivity, and the anomalous Hall conductivity is negligible and shows weak temperature dependence. These observations clearly indicate that the origin of the AHE in $Fe_3Se_4$ is related to sample magnetization which in turn is related to intrinsic Berry phase and/or side-jump scattering.

**Acknowledgements:** The present work has received funding from the Academy of Finland (No. 292431) and made use of the RawMatTERS Finland infrastructure (RAMI) facilities based at Aalto University. DS acknowledges the CSC the Finnish IT Center for Science for computational resources. Dr. T. Björkman is acknowledged for his comments and suggestions on the DFT calculations.